\documentclass[aps,prb,twocolumn,showpacs,groupedaddress]{revtex4}
\usepackage{here}
\usepackage{graphicx}
\begin{document}

% Use the \preprint command to place your local institutional report
% number in the upper righthand corner of the title page in preprint mode.
% Multiple \preprint commands are allowed.
% Use the 'preprintnumbers' class option to override journal defaults
% to display numbers if necessary
%\preprint{}

\title{High-field magnetization study of the S = 1/2 antiferromagnetic
  Heisenberg chain [PM Cu(NO$_3$)$_2$(H$_2$O)$_2$]$_n$ with a field-induced gap}

\author{A.U.B. Wolter$^1$, H. Rakoto$^2$, M. Costes$^2$, A. Honecker$^3$,
  W. Brenig$^3$, A. Kl\"{u}mper$^4$,
  H.-H. Klauss$^1$, F.J. Litterst$^1$, R. Feyerherm$^5$, D. J\'{e}rome$^6$, S. S\"{u}llow$^{1}$}
%\email[]{Your e-mail address}
%\thanks{}
%\altaffiliation{}
\address{$^1$Institut f\"{u}r Metallphysik und Nukleare Festk\"{o}rperphysik,
  TU Braunschweig, 38106 Braunschweig, Germany\\
$^2$Laboratoire National des Champs Magn\'{e}tiques Puls\'{e}s, 31432 Toulouse Cedex 04, France\\
$^3$Institut f\"{u}r Theoretische Physik, TU Braunschweig, 38106 Braunschweig,
  Germany\\
$^4$Fachbereich Physik, Bergische Universit\"{a}t Wuppertal, 42097 Wuppertal, Germany\\
$^5$Hahn--Meitner--Institut GmbH, 14109 Berlin, Germany\\
$^6$Laboratoire de Physique des Solides, Universit\'{e} Paris-Sud, 91405 Orsay
  Cedex, France}

\date{\today}

\begin{abstract}
We present a high-field magnetization study of the $S$=1/2
antiferromagnetic Heisenberg chain [PM
Cu(NO$_3$)$_2$(H$_2$O)$_2$]$_n$. For this material, as result of
the Dzyaloshinskii-Moriya interaction and a staggered $g$-tensor,
the ground state is characterized by an anisotropic field-induced
spin excitation gap and a staggered magnetization. Our data reveal
the qualitatively different behavior in the directions of maximum
and zero spin excitation gap. The data are analyzed via exact
diagonalization of a linear spin chain with up to 20 sites and on
basis of the Bethe ansatz equations, respectively. For both
directions we find very good agreement between experimental data
and theoretical calculations. We extract the magnetic coupling
strength $J/k_B$ along the chain direction to 36.3(5) K and
determine the field dependence of the staggered magnetization
component $m_s$.
\end{abstract}

% insert suggested PACS numbers in braces on next line
% insert suggested keywords - APS authors don't need to do this
%\keywords{}

%\maketitle must follow title, authors, abstract, \pacs, and \keywords
\maketitle

Motivated by the rich variety of different magnetic ground states,
such as quantum critical behavior or gaps in the spin excitation
spectra, quasi-one-dimensional quantum magnets have been the focus
of intense experimental and theoretical research efforts in recent
years ~\cite{bofish,haldane,dender,stone}. To gain deeper insight
into the physics of such quantum spin systems well-defined model
compounds need to be explored. Here, the uniform $S$ = 1/2
antiferromagnetic Heisenberg chain (AFHC) is of particular
interest, since it is exactly solvable using the so-called Bethe
ansatz equations ~\cite{bethe,fledderjohann,klumper2}.

In $S$ = 1/2 AFHC´s, lacking inversion symmetry, additional terms
in the Hamiltonian have to be taken into account, that is the
Dzyaloshinskii-Moriya (DM) interaction and an alternating $g$
tensor ~\cite{AO,oshi2}. This gives rise to an effective staggered
field $h_s$ perpendicular to the applied magnetic field $H$. Then
the Hamiltonian is written as ~\cite{AO,oshi2}

\begin{equation}
\hat{H} = J\sum_{i} [{\bf S}_i{\bf
S}_{i+1}-h_uS_i^z-(-1)^ih_sS_i^x] \label{Hop}
\end{equation}

\noindent with $J$ as the coupling constant, $h_u = g\mu_BH/J$ as
the effective uniform field, and $h_s$ the induced effective
staggered field. In the following we refer to this as the {\it
staggered} $S$ = 1/2 AFHC model. Resulting from this extension of
the uniform $S$ = 1/2 AFHC are the opening of an anisotropic spin
excitation gap with application of a magnetic field and new,
particlelike excitations such as solitons, antisolitons, and their
bound state, the "breather" ~\cite{asano,essler}. Moreover, by
fully evaluating the effect of the DM interaction on the ground
state properties, a crossover to a qualitatively different
high-field behavior has been predicted recently ~\cite{zhao}.

The model for the staggered $S$ = 1/2 AFHC has been used to
describe two materials in particular, copper
benzoate~\cite{dender} and copper pyrimidine nitrate [PM
Cu(NO$_3$)$_2$(H$_2$O)$_2$]$_n$ ~\cite{feyerherm1}. For the latter
compound, from a single-crystal study a magnetic exchange
parameter $J/k_B$ = 36 K is derived. Further, an additional
Curie-like contribution to the magnetic susceptibility at low
temperatures is observed, which varies strongly with magnitude and
direction of the applied external field. Specific-heat
measurements in magnetic fields verify the predicted formation of
an anisotropic spin excitation gap, whose magnitude also depends
on size and orientation of the external field ~\cite{feyerherm1}.
The spin excitation gap and the Curie-like contribution to $\chi$
are largest for the same field direction (referred to as $c''$,
for notation see Ref. 13) and vanish for one direction
perpendicular to $c''$ in the $a-c$ plane (referred to as $a''$).

For $S$ = 1/2 AFHC materials, (high-field) magnetization
experiments are abundant and are perfectly described by theory
~\cite{bofish,griffiths,mollymoto,azevedo,hammar,klumper}.
Recently, for the staggered $S$ = 1/2 AFHC the magnetization curve
has been calculated by several groups ~\cite{zhao,ueda,lou,gross}.
As yet, these theoretical predictions have not been verified
experimentally.

Therefore, in this Rapid Communication we present a magnetization
study on [PM Cu(NO$_3$)$_2$(H$_2$O)$_2$]$_n$ covering the entire
field range up to saturation, {\it i.e.} $\mu_0H$ = 53 T. With our
study of the magnetization along $a''$ and $c''$ we establish the
contrasting behavior along these two directions, the first
representing the uniform, and the latter the staggered $S$ = 1/2
AFHC. The behavior of the uniform $S$ = 1/2 AFHC is evaluated on
basis of the Bethe ansatz equations ~\cite{klumper2,klumper}. In
contrast, for the staggered $S$ = 1/2 AFHC we analyze our data by
means of exact diagonalization of linear chains with up to $N$ =
20 spins, based upon the staggered field theory by Oshikawa and
Affleck ~\cite{AO,oshi2}. From this analysis we find very good
agreement between experimental data and theoretical calculations.
We determine the characteristic parameters, {\it i.e.}, the
coupling constant $J/k_B$ and the staggered magnetization $m_s$.

Comparing our finite-size calculations to previous density-matrix
renormalization group (DMRG) studies ~\cite{zhao,ueda,lou,gross}
for the Hamiltonian (\ref{Hop}), we find perfect agreement between
our results and those of other groups. Only for low-fields
finite-size effects are present. The advantage of our finite-size
calculations compared to the DMRG method lies in its simplicity
and short computation time.

Single crystals of [PM Cu(NO$_3$)$_2$(H$_2$O)$_2$]$_n$ have been
grown by slow evaporation of the equimolar aqueous solution of
copper nitrate and pyrimidine ~\cite{ishida}. The crystals show
well-defined facets and the principal axes can be identified
easily. We have checked by low-field magnetization measurements
that the magnetic susceptibility matches the one published in Ref.
13. For the magnetization measurements the samples were oriented
along the characteristic orientations $a''$ and $c''$
(misalignment $\le$ 5$^o$), glued to the tip of a plexi glass rod
and placed inside a thin walled teflon cylinder. The magnetization
signal of the sample holder was negligible.

Magnetization measurements were carried out at the Laboratoire
National des Champs Magn\'{e}tiques Puls\'{e}s in Toulouse in
pulsed magnetic fields up to $\mu_0H$ = 53 T. Pulsed magnetic
fields were obtained by discharging a capacitor bank in a solenoid
according to a crowbar described in Ref. 23. The pulse duration
was about 200 ms with an increasing time of 25 ms. The
magnetization was detected as a voltage $V$ induced in a
compensated arrangement of pick-up coils wound concentrically
around the sample and coupled to it with the coupling constant
$\eta$, such that $V = \eta \Omega \delta M/\delta t$, with
$\Omega$ as the sample volume ~\cite{note1}. The absolute
magnetization was obtained by numerical integration of this
voltage. Due to the limited sample space (diameter $<$ 1.6 mm),
the absolute signal was small ($<$ 10$^{-5}$ Am$^2$). To achieve a
higher accuracy of the signal calibration additional measurements
were performed in magnetic fields up to 5 T in a commercial
superconducting quantum interference device magnetometer.

\begin{figure}
\begin{center}
\includegraphics[width=0.9\columnwidth]{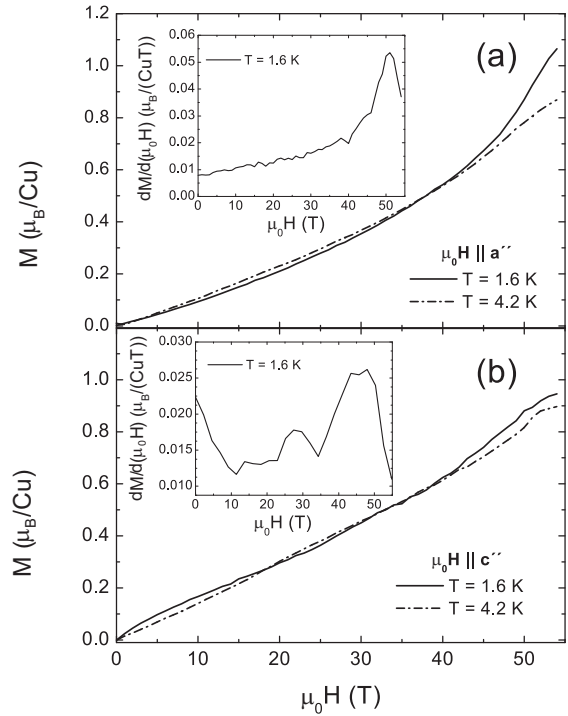}
\end{center}
\caption[1]{The field dependence of the magnetization of [PM
  Cu(NO$_3$)$_2$(H$_2$O)$_2$]$_n$ with the external field aligned along the
  $a''$ (a) and the $c''$ direction (b). In the insets, the field derivative
  $dM/d(\mu_0H)$ is displayed.} \label{fig:fig1}
\end{figure}

In Fig. 1 we present the magnetization curve of [PM
Cu(NO$_3$)$_2$(H$_2$O)$_2$]$_n$ as a function of field at 1.6 and
4.2 K for the two characteristic orientations, i.e., $H || a''$
(Fig. 1(a)) and $|| c''$ (Fig. 1(b)), respectively. Comparing the
magnetization along the two directions, an anisotropic response is
observed. At 1.6 K for $H || a''$ we find the archetypical
behavior of the $S$ = 1/2 AFHC ~\cite{hammar}. In contrast, for
fields parallel to the $c''$ axis an additional low field
contribution and a delayed saturation of the magnetization occurs.
To emphasize this difference in the insets of Figs. 1(a) and (b)
we plot the derivatives of the magnetization, $dM/d(\mu_0H)$. For
small fields $|| c''$ the initial slope is more than twice as
large as for the $a''$ axis. At high fields ($>$ 35 T), the
saturation of the magnetization for $H || c''$ is suppressed, as
indicated by a smaller, broader feature in $dM/d(\mu_0H)$, as
compared to the $a''$ axis response. Increasing the temperature to
4.2 K reduces the difference in the $M (\mu_0H$) curves between
the two directions, but does not completely suppress it.

The deviation from the uniform $S$ = 1/2 AFHC behavior along the
$c''$ direction is attributed to an additional magnetization
component. It increases much faster and passes through a maximum
at a much lower field than the uniform saturation field $\mu_0H
_{sat}$. Since the second component is not present along $a''$, we
ascribe it to the staggered magnetization $m_s$.

The $g$ tensor of [PM Cu(NO$_3$)$_2$(H$_2$O)$_2$]$_n$ has been
derived from electron spin resonance measurements
~\cite{feyerherm1}. In the uniform $S$ = 1/2 AFHC model the
saturation field is calculated according to the formula
\cite{oshi2}

\begin{equation}
H_c = 4JS/g\mu_B.
\end{equation}

\noindent For zero temperature and using
$J/k_B$\,=\,36\,$\pm$\,0.5\,K, $g_{a''}$\,=\,2.14\,$\pm$\,0.02,
and $g_{c''}$\,=\,2.21\,$\pm$\,0.02 from Ref. 13, we obtain $\mu_0
H_{sat}$\,=\,50.1\,$\pm$\,0.8\,T and 48.5\,$\pm$\,0.8\,T along
$a''$ and $c''$ axes, respectively. The saturation magnetization
$M_s$ is calculated to
$M_{s,a''}$\,=\,1.07\,$\pm$\,0.01\,$\mu_B$/Cu atom and
$M_{s,c''}$\,=\,1.11\,$\pm$\,0.01\,$\mu_B$/Cu atom. Thus, for the
uniform $S$ = 1/2 AFHC at $T$ = 1.6 K $\ll$ $J/k_B$, the
saturation magnetization should be approached at highest
experimental applied field. Indeed for $H || a''$ (Fig. 1(a)), the
$T$ = 1.6 K curve has an initial slope lower than at 4.2 K. With
increasing field the curvature becomes larger, crosses the 4.2 K
curve near 38 T and almost reaches saturation at $\approx$ 53 T.
The data for both temperatures become nonlinear with field for
$\mu_0H >$ 15 T. Moreover, with decreasing temperature the data
sets approach the $T$ = 0 curve for the uniform $S$ = 1/2 AFHC in
full agreement with previous experimental work ~\cite{hammar}.

\begin{figure}
\begin{center}
\includegraphics[width=1\columnwidth]{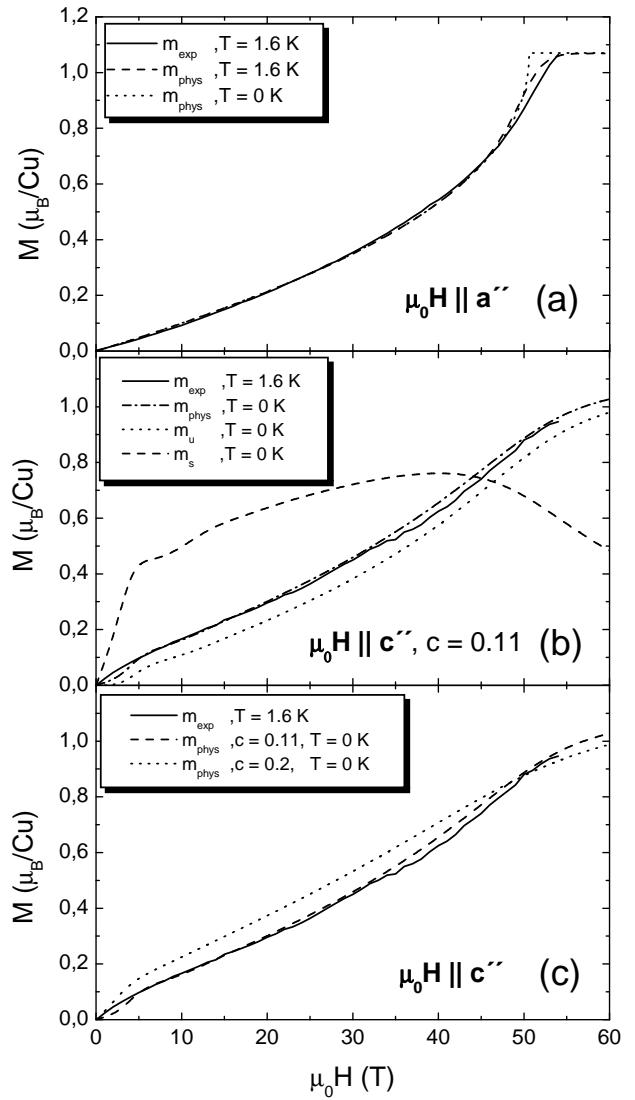}
\end{center}
\caption[2]{Experimental and theoretical magnetization curves for [PM
  Cu(NO$_3$)$_2$(H$_2$O)$_2$]$_n$ at $T$ = 1.6 K for different field
  directions. (a) $\mu_0H || a''$: solid line - experimental data, dashed line
  - fit assuming uniform $S$ = 1/2 AFHC for $T$ = 1.6\,K, dotted line - $T$ = 0 calculations ~\cite{klumper}. (b) $\mu_0H || c''$:
  dotted line - uniform magnetization $m_u$, dashed line - staggered
  magnetization $m_s$, dash-dotted line - calculated physical magnetization
  $m_{phys}$ for c =
  0.11, solid line - experimental data. (c) $\mu_0H || c''$: dashed line -
  $m_{phys}$ for $c$ = 0.11, dotted line - $m_{phys}$ for $c$ = 0.2, solid line - experimental data.}
\end{figure}

From a theoretical point of view, the magnetization curve of the
uniform $S$ = 1/2 AFHC has been computed for $T = 0$ by Bethe
ansatz ~\cite{griffiths}. However, our measurements have been
carried out at temperatures $T > 0.04 J$, where thermal
fluctuations lead to a substantial rounding of the cusp at the
saturation field in the $T = 0$ magnetization curve. On the other
hand, at 1.6\,K we have $T < 0.05J$, substantially smaller than
the lowest corresponding temperature studied in Ref. 17 for copper
pyrazine dinitrate. Even at $T = 0.05J$, the magnetization curve
computed for the uniform $S$=1/2 AFHC on a ring with $N=16$ sites
still exhibits clear finite-size effects, precluding an analysis
along the lines of Ref. 17. We therefore use results obtained by
the thermodynamic Bethe ansatz in the thermodynamic limit $N =
\infty$ and at arbitrary temperature $T$ to describe the
magnetization curve for $H\parallel a''$ ~\cite{klumper}.

The situation for $H\parallel c''$ is quite different. On the one
hand, the Hamiltonian (\ref{Hop}) cannot be solved exactly with a
nonzero staggered field $h_s$ and we therefore have to rely on a
numerical treatment. Whereas the $z$-component $S^z$ of the total
spin is conserved for $h_s=0$, even this is not the case anymore
for $h_s\ne 0$. This has two consequences: ({\it i}) The reduced
symmetry of the Hamiltonian (\ref{Hop}) restricts the system sizes
$N$ that can be accessed, and ({\it ii}) each new set of magnetic
fields $h_u$ and $h_s$ requires a new numerical determination of
the ground state. On the other hand, the field-induced opening of
a gap ~\cite{AO,oshi2} leads to the following two simplifications:
({\it i}) For the high magnetic fields studied here, the gap is
sufficiently large to suppress thermal excitation at low
temperatures. Therefore, finite temperature is expected to have
only a comparatively small effect. Indeed, this is confirmed by
the difference of the $T = 1.6$\,K and $4.2$\,K curves in Fig.\ 1,
which is noticeably smaller along the $c''$ direction (panel (b))
than for the $a''$ one (panel (a)). This permits us to compare a
measurement at low but finite $T$ with a computation at $T = 0$.
({\it ii}) In most field ranges considered here, the correlation
length turns out to be sufficiently short such that finite-size
effects can be neglected already for systems with only $N = 20$
sites. More precisely, the correlation length is large only for a
small staggered field $h_s$ and thus only the low-field region
suffers from finite-size effects. These lead to an artificial
low-field peak in the staggered magnetization $m_s$ (see e.g.,\
dashed curve in Fig.\ 2(b)) whose position roughly determines the
region up to which finite-size effects are still relevant, as
evidenced by comparison with results for $N \le 16$ (not shown).
Due to the fast disappearance of finite-size effects for higher
magnetic fields it is completely sufficient for our purposes to
apply the Lanczos diagonalization procedure to rings with $N\le20$
sites. Therefore, the additional effort of a DMRG procedure
~\cite{ueda,lou,gross,zhao} is not necessary here.

A final remark is in order before we present our numerical results
and compare them with the experiment. In the presence of a
staggered field $h_s = c \, h_u$, which is related by a constant
anisotropy parameter $c$ to the uniform field $h_u$, the physical
magnetization $m_{phys}$ (which is measured experimentally) is
given by the superposition of the uniform $m_u$ and staggered
$m_s$ magnetization components ~\cite{AO,oshi2}:

\begin{equation}
m_{phys} = m_u + c \, m_s.
\label{mphys}
\end{equation}
To our knowledge, previous numerical works ~\cite{ueda,lou,gross,zhao} only
showed $m_u$ and $m_s$ separately, while numerical results for the combination
$m_{phys}$ (Eq. \ref{mphys}) have not been published yet.

For the magnetization along $a''$ (Fig. 2(a)), using
$\mu_0H_{sat}$ = 50.6\,T, $g_{a''}$\,=\,2.14, and taking into
account the finite experimental temperature, we find very good
agreement between our data and the Bethe ansatz result
~\cite{klumper}. The deviations for fields $\mu_0H >$ 50 T can be
attributed to the misalignment of the crystal, while for smaller
fields experimental data and calculated result match within
2\,$\%$. Thus, for this field direction the uniform $S$ = 1/2 AFHC
is established.

In Fig. 2(b) we depict the field dependence of the magnetization
along $c''$. For fields $\mu_0H >$ 10\,T we can fully describe
these data on basis of the staggered $S$ = 1/2 AFHC model using
$\mu_0H_{sat}$ = 49.3\,T, $g_{c''}$ = 2.19. Here, we have used an
anisotropy parameter $c$ = 0.11. This value has been obtained from
a comparison of the data and calculations for $c$ $\in$
[0.08;0.12], step size 0.01, as the optimum solution. The values
$\mu_0H$ along $a''$ and $c''$ are fully consistent with those
predicted from Eq. (2) within their error bars. On average, the
saturation fields $||$ $c''$ and $||$ $a''$ correspond to a
magnetic coupling strength $J/k_B$ = 36.3(5) K.

With the value of the anisotropy parameter $c$ we can decompose
$m_{phys}$ into the uniform ($m_u$) and staggered ($m_s$)
component, according to Eq. (\ref{mphys}). Both, $m_u$ and $m_s$,
are included in Fig. 2(b). Their field dependence closely
resembles the ones obtained in Ref. 20 for the case $h_u$ =
10\,$h_s$. Specifically, we find that $m_s$ traverses a maximum at
$\sim$ 40\,T, while $m_u$ and $m_s$ approach finite but
nonsaturated values for largest fields. Our analysis establishes
the staggered $S$ = 1/2 AFHC for the $c''$ axis of [PM
Cu(NO$_3$)$_2$(H$_2$O)$_2$]$_n$.

Further, we performed calculations for $c$ values in the range
0.08\,-\,0.28. The cases $c$\,= \,0.2 and 0.11 are depicted in
Fig. 2(c). From these data it appears that for decreasing
parameter $c$ the curvature of $m_{phys}$ increases and approaches
the uniform $S$ = 1/2 AFHC magnetization behavior for vanishing
staggered field $h_s$, as expected. With the definition of the
Hamiltonian of the staggered $S$ = 1/2 AFHC model in Ref. 13 the
anisotropy parameter $c$\,=\,0.11 from this work translates into a
value 0.24 ~\cite{note}. Thus, we find perfect agreement of our
$c$ value with the one obtained from magnetic susceptibility
measurements in Ref. 13, {\it i.e.}, $c$\,=\,0.235.

In conclusion, we have performed high-field magnetization
experiments on [PM$\cdot$Cu(NO$_3$)$_2$$\cdot$(H$_2$O)$_2$]$_n$ at
temperatures 1.6 and 4.2\,K. We analyzed our data by means of
Bethe ansatz equations for the experiments along the $a''$ axis
and exact diagonalization of linear chains containing $N$ = 20
spins, based upon the staggered field theory by Oshikawa and
Affleck ~\cite{AO,oshi2}, for the $c''$ axis. The very good
agreement of our data with our theoretical calculations and those
of other groups ~\cite{ueda,lou} verifies the predictions for the
uniform and staggered $S$ = 1/2 AFHC models. This way, for the
staggered case we have extracted the staggered magnetization
component $m_s$\,($\mu_0H$).

Recently, for the staggered $S$ = 1/2 AFHC model a
low-field/high-field crossover in the staggered magnetization has
been predicted ~\cite{zhao}. For
[PM$\cdot$Cu(NO$_3$)$_2\cdot$(H$_2$O)$_2$]$_n$ these calculations
would imply deviations from the staggered behavior in the field
range $\mu_0H >$ 50\,T, and thus cannot be observed in our study.
To verify the predictions from Ref. 12, in the future it will be
interesting to perform analogous magnetization studies on related
materials with smaller $J$ values.

This work has partially been supported by funds of the European contract
no. HPRI-CT-1999-40013 and by the DFG under contract no. SU229/6-1.


\begin{references}
\bibitem{bofish} J.C. Bonner and M.E. Fisher, Phys. Rev. {\bf 135}, A640 (1964).
\bibitem{haldane} F.D.M. Haldane, Phys. Rev. Lett. {\bf 50}, 1153
(1983).
\bibitem{dender} D.C. Dender, P.R. Hammar, D.H. Reich, C. Broholm, and
  G. Aeppli, Phys. Rev. Lett. {\bf 79}, 1750 (1997).
\bibitem{stone} M.B. Stone, D.H. Reich, C. Broholm, K. Lefmann, C. Rischel,
  C.P. Landee, and M.M. Turnbull, Phys. Rev. Lett. {\bf 91}, 037205 (2003).
\bibitem{bethe} H.A. Bethe, Z. Phys. {\bf 71}, 205 (1931).
\bibitem{fledderjohann} A. Fledderjohann, C. Gerhardt, K.H. M\"{u}tter,
  A. Schmitt, and M. Karbach, Phys. Rev. B {\bf 54}, 7168 (1996).
\bibitem{klumper2} A. Kl\"{u}mper and D.C. Johnston, Phys. Rev. Lett. {\bf
    84}, 4701 (2000).
\bibitem{AO} M. Oshikawa and I. Affleck, Phys.\ Rev.\ Lett.\ {\bf 79}, 2883 (1997).
\bibitem{oshi2} I. Affleck and M. Oshikawa, Phys. Rev. B, {\bf 60}, 1038
  (1999).
\bibitem{asano} T. Asano, H. Nojiri, Y. Inagaki, J.P. Boucher, T. Sakon,
  Y. Ajiro, and M. Motokawa, Phys. Rev. Lett. {\bf 84}, 5880 (2000).
\bibitem{essler} F.H.L. Essler and A.M. Tsvelik, Phys. Rev. B {\bf 57}, 10592
  (1998).
\bibitem{zhao} J.Z. Zhao, X.Q. Wang, T. Xiang, Z.B. Su, and L. Yu,
  Phys. Rev. Lett. {\bf 90}, 207204 (2003).
\bibitem{feyerherm1} R. Feyerherm, S. Abens, D. G\"{u}nther, T. Ishida,
  M. Mei\ss ner, M. Meschke, T. Nogami, and M. Steiner, J. Phys.:
  Condens. Matter {\bf 12}, 8495 (2000).
\bibitem{griffiths} R.B.\ Griffiths, Phys. Rev. {\bf 133}, A768 (1964).
\bibitem{mollymoto} H. Mollymoto, E. Fujiwara, M. Motokawa, and M. Date,
J. Phys. Soc. Jpn. {\bf 48}, 1771 (1980).
\bibitem{azevedo} L.J. Azevedo, A. Narath, P.M. Richards, and Z.G. Soos,
  Phys. Rev. B {\bf 21}, 2871 (1980).
\bibitem{hammar} P.R. Hammar, M.B. Stone, D.H. Reich, C. Broholm, P.J. Gibson,
  M.M. Turnbull, C.P. Landee, and M. Oshikawa, Phys. Rev. B {\bf 59}, 1008
  (1999).
\bibitem{klumper} A.\ Kl\"umper, Eur.\ Phys.\ J.\ B {\bf 5}, 677 (1998).
\bibitem{ueda} N. Shibata and K. Ueda, J. Phys. Soc. Jpn. {\bf 70}, 3690
  (2001).
\bibitem{lou} J. Lou, S. Qin, C. Chen, Z. Su, and L. Yu, Phys. Rev. B {\bf 65},
  064420 (2002).
\bibitem{gross} F.\ Capraro and C.\ Gros, Eur.\ Phys.\ J.\ B {\bf 29}, 35 (2002).
\bibitem{ishida} T. Ishida, K. Nakayama, M. Nakagawa, W. Sato, Y. Ishikawa,
  M. Yasuri, F. Iwasaki, and T. Nogami, Synth. Met. {\bf 85}, 1655 (1997).
\bibitem{pulse} O. Portugall, F. Lecouturier, J. Marquez, D. Givord, and S. Ask\'{e}nazy, Physica B {\bf 294-295}, 579 (2001).
\bibitem{note1} Measurements were made for increasing and decreasing field. No hysteresis has been observed. For clarity,
because of a higher signal-to-noise ratio, in the figures we only present data taken for decreasing field.
\bibitem{note} Due to different prefactors in the Hamiltonian from
  Ref. 13 and in Eq.(\ref{Hop}) of this work the $g$ factor has to be
  multiplied with our $c$ value to obtain the anisotropy parameter from Ref. 13.
\end{references}
\end{document}